\documentclass[12pt]{article}

\usepackage[margin=1in,paper=letterpaper]{geometry}
\usepackage{setspace}
\usepackage{times}
\usepackage{booktabs}
\usepackage{tabularx}
\usepackage{array}
\usepackage{graphicx}
\usepackage[hidelinks]{hyperref}
\usepackage{natbib}
\usepackage{caption}
\usepackage{float}
\usepackage{amsmath}
\usepackage{enumitem}
\usepackage{etoolbox}

\AtBeginEnvironment{table}{\singlespacing}

\newcommand{\citepos}[1]{\citeauthor{#1}'s (\citeyear{#1})}

\newcommand{\AII}{AI-interviewer}
\newcommand{\AIIs}{AI-interviewers}

\bibpunct{(}{)}{;}{a}{}{: }

\begin{document}

\begin{titlepage}
  \centering
  \vspace*{2cm}
  {\LARGE\bfseries Can AI interview? Comparing AI-led to Human-led
    Chat Interviews\par}
  \vspace{1.5cm}
  {\large
    Semra Guven\textsuperscript{1,*} \quad
    Tobias G{\aa}rdhus\textsuperscript{2,*} \quad
    Andreas Bjerre-Nielsen\textsuperscript{2} \\[0.4em]
    Anna Rogers\textsuperscript{3} \quad
    Hjalmar Bang Carlsen\textsuperscript{2,\dag}\par}
  \vspace{1em}
  {\small
    \textsuperscript{1}University of Copenhagen, Denmark\\
    \textsuperscript{2}Center for Social Data Science, University of Copenhagen, Denmark\\
    \textsuperscript{3}Data Science, IT University of Copenhagen, Denmark\\[0.6em]
    \textsuperscript{*}These authors contributed equally to this work.\\
    \textsuperscript{\dag}Corresponding author: \href{mailto:hc@sodas.ku.dk}{hc@sodas.ku.dk}\par}
  \vspace{1cm}
  {\large October 2026\par}
  \vspace{2cm}
  \textbf{Keywords:} AI interviews; qualitative methods; large
  language models; data quality; open-source models
\end{titlepage}

% -------------------------------------------------------
\begin{abstract}
\noindent
Large language model chatbots have been proposed as \AIIs{} for collecting qualitative interview data at scale, potentially bridging social science's longstanding divide between the breadth of surveys and the depth of qualitative interviews \citep{lazarsfeld_controversy_1944}. This paper provides the first direct empirical comparison of data quality in AI-led and human-led interviews. In a randomized experiment ($N = 41$), university students were interviewed via synchronous text chat by either trained non-expert interviewers or a locally hosted AI system; we evaluate exposure, specificity, and relevance. Median interview duration was similar across conditions, but \AIIs{} asked more than twice as many questions per interview and elicited roughly 65 percent more respondent text per interview. Per-question responses were longer with human interviewers, though not after multiple-testing correction; only AI-condition participants skipped questions. Responses were almost uniformly on-topic in both conditions; for specificity we detect no difference, but the design cannot exclude one of up to three quarters of a point. These results were obtained with a small, locally hosted open-weight model (7B parameters), which addresses concerns about data security, GDPR compliance, and scientific reproducibility. Our findings support the viability of AI-interviewing for large-$N$ qualitative data collection.
\end{abstract}

\newpage

% -------------------------------------------------------
\section*{Introduction}

\AIIs{} powered by large language models (LLMs) have recently been proposed as a solution to a longstanding methodological challenge in sociological research: how to collect qualitative interview data at scale \citep{Chopra2023, geiecke_conversations_2024, Wuttke2024, jacobsen_chatbots_2025}. \AIIs{} typically ask open-ended questions and generate contextually responsive follow-up probes, enabling respondents to provide detailed answers argued to yield higher-quality data than traditional open-ended survey instruments \citep{geiecke_conversations_2024}. By bridging the breadth of standardized surveys with the depth and flexibility of semi-structured interviews, \AIIs{} seek to address what \citet{lazarsfeld_controversy_1944} already identified as a fundamental challenge in social research: reconciling the scale of survey methods with the depth and detail of qualitative interviewing.

The appeal of \AIIs{} rests on a structural constraint in qualitative research. Qualitative approaches yield accounts of lived experience and social process that surveys cannot capture, but the costs of data collection have traditionally limited sample sizes, thereby hindering population-level inference \citep{small_how_2009, Goldthorpe2000, Weiss1995}. \AIIs{} seek to address this constraint by radically reducing the cost of qualitative data collection, enabling interview-based research at scales previously feasible only with survey instruments \citep{geiecke_conversations_2024, Wuttke2024}.

A critical prerequisite for this promise, however, is that AI-led interviews produce data of sufficient quality. If the data generated by \AIIs{} are substantially weaker than those produced by human interviewers, their methodological utility is limited. To assess this, \AIIs{} must be evaluated against established standards of qualitative evidence \citep{SmallCalarco2022}. Yet to date, no systematic empirical study has directly compared data quality across AI-led and human-led interviews, leaving a critical gap in the methodological literature.

We address this gap with a randomized experiment in which $N = 41$ participants were assigned to either a human-led or AI-led synchronous chat interview. Both conditions used the same chat interface, topic, and interview guide. We evaluate data quality across three dimensions widely used in qualitative methods research: exposure, specificity, and relevance \citep{SmallCalarco2022, Weiss1995, Lareau2021}. Our findings show that responses were overwhelmingly on-topic in both conditions and reveal no detectable difference in specificity. Median interview duration is similar across conditions, but \AIIs{} ask more than twice as many questions per minute and collect substantially more respondent text per minute. We conclude by discussing the challenges inherent in quantitatively evaluating qualitative data quality and the implications for the wider development of AI-interviewing as a sociological method.

Beyond the quality comparison itself, this study makes a methodological contribution in a second respect: we implement the \AII{} using a locally hosted, open-weight language model rather than a proprietary API system. To the best of our knowledge, ours is the first \AII{} solution based on a relatively small (7B parameters) open-weight model.\footnote{An early version of this work was released as a preprint in July 2025. In November 2025, \citet{geiecke_conversations_2024} updated their preprint with a comparison involving two Llama models, both substantially larger than the model used in this study; that study also does not compare AI-led interviews with human-led interviews.} Existing AI-interview research has relied almost exclusively on large commercial models, chiefly variants of ChatGPT --- a practice that raises serious concerns about data security, GDPR compliance, and scientific reproducibility \citep{palmer_using_2024, Spirling2023}. Our implementation demonstrates that a 7B-parameter open-weight model can deliver acceptable interview performance while addressing these concerns. We see the development of open-source AI-interview infrastructure as a pressing methodological priority for the discipline \citep[see also][]{palmer_using_2024, Spirling2023}.

% -------------------------------------------------------
\section*{System Design: Local, Open, and Restricted}

As shown in Table~\ref{tab:prior}, most existing AI-interview research relies on commercial variants of ChatGPT --- large, proprietary, API-based language models. Our approach differs in two key respects: we use a smaller, locally hosted, open-weight model, and we deliberately constrain the degrees of freedom granted to the \AII{}. This section explains the rationale behind these design choices and describes the resulting system.

\begin{table}[H]
\centering
\caption{Overview of AI-Interviewer System Implementations}
\label{tab:prior}
\begin{tabularx}{\textwidth}{lX}
\toprule
\textbf{Study} & \textbf{Model used} \\
\midrule
\citet{Chopra2023}     & GPT-4-0613, GPT-4-0314 \\
\citet{geiecke_conversations_2024} & GPT-4o-2024-05-13, Llama 3.1 405B, Llama 4 Maverick 400B \\
\citet{Wuttke2024}             & GPT-4-turbo-2024-04-09 \\
\citet{yu_modular_2025}            & GPT-4, GPT-4o, Claude 3 Sonnet \\
\citet{jacobsen_chatbots_2025}     & GPT-4o-2024-05-13 \\
This study                         & OpenHermes 2.5 (Mistral 7B, locally
                                     hosted) \\
\bottomrule
\end{tabularx}
\end{table}

\subsection*{The Case for Local, Open-Weight Models}

We have chosen to develop our own AI-interviewer system, which enables the use of local and open-weight models. This decision is motivated by three factors, which we detail below.

First, local models allow for \textit{data security, control and confidentiality}. API-based models require researchers to transmit interview data to an external third party. Qualitative interviews are typically conducted under conditions of confidentiality, where participants share personal narratives that may be sensitive. Locally hosted models, run on researcher-controlled servers, give the researcher full control over the data, its use, and access \citep{kaiser_protecting_2009}. Second, local models ease \textit{legal compliance}. Under the General Data Protection Regulation (GDPR), transmitting such data to a commercial provider necessitates a formal data processing agreement and assurances of compliance --- requirements that are difficult to guarantee with large commercial providers. Third, local models increase \textit{reproducibility and reliability}. API-based models are updated without announcement, may incorporate undisclosed system prompts or parameters, and can be deprecated at any time. As \citet{palmer_using_2024} and \citet{Spirling2023} have argued, this opacity undermines the reproducibility and reliability of research findings. One part of the interview sample can then differ systematically from the other due to a change in the AI-interviewer provoked by unknown changes to the underlying model. A locally hosted model can be shared and re-used, enabling precise replication across interview rounds, also allowing other researchers to replicate a given study design in other settings.

\subsection*{System Architecture}

\begin{figure}[H]
  \centering
  \includegraphics[width=\linewidth,height=0.72\textheight,keepaspectratio]{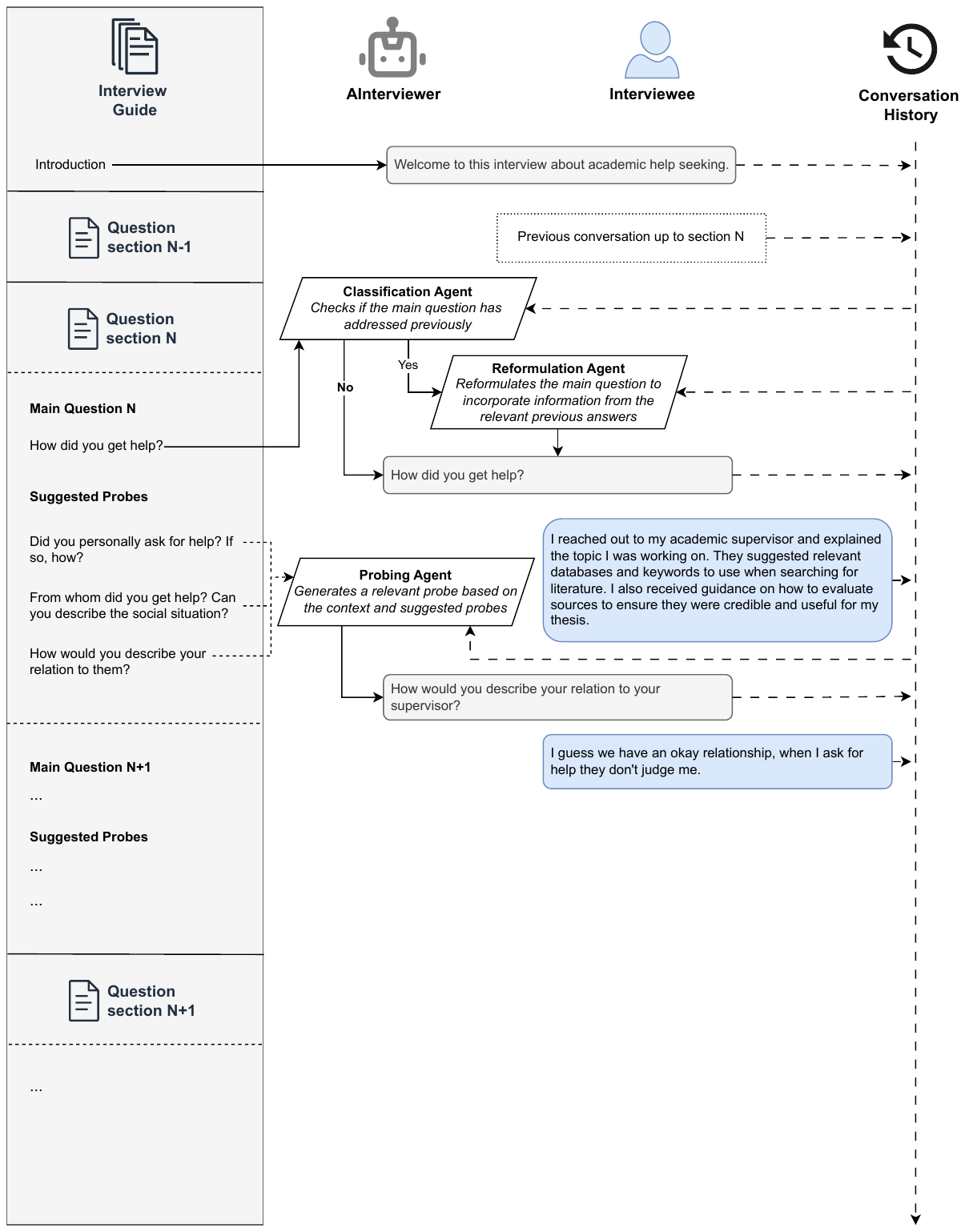}
  \caption{Proposed implementation of the \AII{}. The system is provided with the interview guide and a set of predefined main questions and suggested probes. Three agents --- classification, reformulation, and probing --- coordinate to generate contextually responsive follow-up questions while avoiding duplication.}
  \label{fig:implementation}
\end{figure}

Our \AII{} system follows a multi-agent design inspired by \citet{Chopra2023}, in which distinct LLM components are each assigned a specific task (see Figure~\ref{fig:implementation}). The system employs three agents: (1) a \textit{classification agent} that makes the binary decisions controlling interview flow --- whether a screening question was answered affirmatively, so that the section proceeds or is skipped, and whether a main question has already been addressed in prior responses, so that it is not asked again; (2) a \textit{question reformulation agent} that rephrases questions when necessary to incorporate context from earlier answers; and (3) a \textit{probing agent} that generates contextually responsive follow-up questions. The probing agent is the most central component: its capacity to produce relevant, responsive follow-up probes is the key advantage of AI-interviewing over static survey instruments.

Our system restricts the \AII{} to generating follow-up probes in response to a predefined set of open-ended main questions. Our design thereby differs from earlier approaches in which the AI system freely generates new questions under broad thematic headings \citep{Chopra2023, geiecke_conversations_2024}. This design deliberately limits the AI's autonomy, ensuring that the interview remains anchored in questions the researcher can justify theoretically, epistemologically, and ethically.

All three agents ran on OpenHermes 2.5, an instruction-tuned version of Mistral 7B \citep{jiang2023mistral7b} (7.24B parameters, in Q4\_0 4-bit quantization; Ollama tag \texttt{openhermes:v2.5}), served locally with Ollama. The probing and reformulation agents sampled at temperature 0.7 and the classification agent at temperature 0. All other sampling parameters were left at Ollama's defaults.

% -------------------------------------------------------
\section*{Experimental Design}

\subsection*{Treatment: AI or Human-Led Chat Interviews}

\begin{figure}[H]
  \centering
  \includegraphics[width=\linewidth]{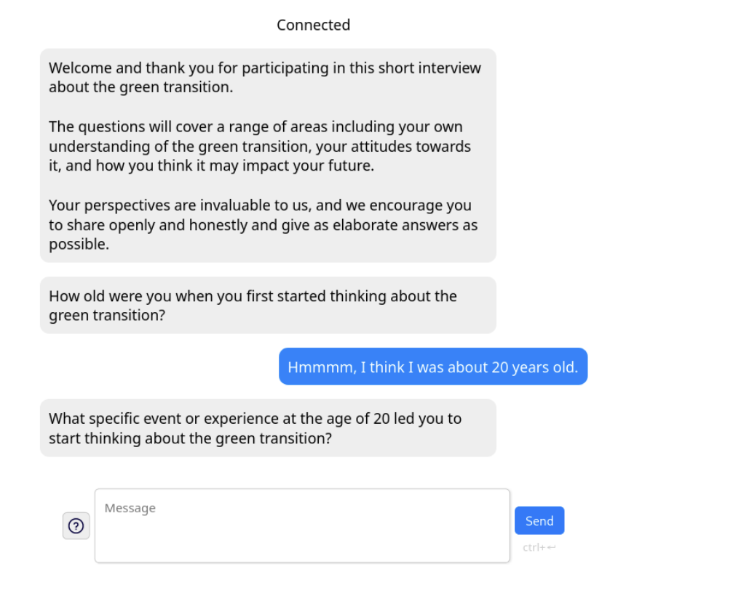}
  \caption{Screenshot of the chat interface used for both human-led and AI-led interviews. The interface displayed conversation history in the upper portion of the screen; a resizable text box at the bottom was used by interviewees to type their responses. The send button was disabled until the next question appeared, prompting participants to submit their response in a single message.}
  \label{fig:screenshot}
\end{figure}

All interviews were conducted through a custom-built chat interface resembling a standard messaging application (see Figure~\ref{fig:screenshot}). The interface displayed conversation history above a resizable text input field; the send button was disabled until the next question appeared, prompting complete single-message responses rather than fragmented entries. A dynamic typing animation simulated real-time response behavior, following recommendations from \citet{Chopra2023} to enhance engagement and reduce attrition.

Each participant was randomly assigned to either a human-led or AI-led interview and was explicitly informed of their assignment upon arrival. Both the AI-interview system and the human interviewers operated under identical instructions, and all interviews were conducted through the same chat interface to hold the interview mode constant \citep{dillman_survey_2005}. The primary source of variation between conditions is the interviewer's identity --- human or AI --- and the relative capacity of each to ask relevant, responsive follow-up questions.

Because participants were informed of their condition before the interview began, we cannot rule out behavioral effects arising from this information. Participants may have responded differently depending on whether they believed they were being interviewed by a human or a machine: they may have disclosed more or less, elaborated at different length, adopted a more or less formal register, or adjusted their effort based on perceived expectations.

\subsection*{Interview Guide and Instructions}

The topic was university students' academic help-seeking practices and experiences \citep{small_someone_2017}. The guide was divided into two main sections: the first addressed a situation in which the student needed and successfully received help; the second addressed a situation in which the student needed help but did not ask for it. Following best practices in qualitative interviewing, the guide was structured to move from concrete experiences to feelings and finally to general reflections \citep{Gerson2020, Weiss1995}. A third case, in which the student asked for help but did not receive it, was administered during the first three days of data collection and then dropped: the sessions were scheduled in half-hour slots and the human interviewers rarely reached it, so retaining it would have meant comparing conditions on material only one of them regularly covered. It was administered in eight of the 41 interviews and is excluded from all analyses reported here.

Each section opened with a yes/no screening question establishing that the respondent had such a situation to describe, followed by six open-ended main questions, each paired with suggested follow-up probes (see Appendix~\ref{app:guide} for the full guide). Both human and AI interviewers were instructed to administer the guide consistently while adapting probes to the specific content of each response. All interviewers were also given general guidelines drawn from established qualitative interviewing practice \citep{Lareau2021, Gerson2020, Weiss1995}: probe for more detail; ask one question at a time; reformulate if misunderstood; refrain from judgment; avoid leading questions; and respect when participants decline to answer.

\subsection*{Interviewers, Setting, and Recruitment}

The two human interviewers had prior experience with qualitative interviewing through advanced methods coursework or research experience. While not expert interviewers, their level of training reflects what is typical in large-scale qualitative data collection projects where non-expert research assistants are trained in a specific guide and topic, as in projects such as the American Voices Project \citep{EdinFieldsEtAl_2024_Listening_to_Voices_of_America}. Both conducted pilot interviews before data collection to refine the guide and their technique.

Forty-one participants (university students) were recruited through face-to-face outreach, flyers, and posters on campus and assigned to fixed time slots. All interviews were conducted on-site: participants and interviewers used their own devices to access the chat interface while seated in adjacent rooms. The participants could not see and did not interact with the interviewers outside the chat during the interviews. This on-site design ensured genuine engagement with the interview, avoiding the growing problem of respondents using generative AI tools to answer questions on online platforms --- a recognized threat to data quality in remote studies \citep{ZhangXuEtAl_2025_Generative_AI_Meets_Open-Ended_Survey_Responses_Research_Participant_Use_of_AI_and_Homogenization}. Participants received a voucher for a free lunch at the university canteen as compensation. All interviews were conducted in English. At the conclusion of each interview, participants completed a short online survey recording their background characteristics (reported in Table~\ref{tab:balance}).

\subsubsection*{Randomization}

Human-led and AI-led interviews were conducted in parallel within each time slot, allowing participants to be randomly assigned to conditions. Randomization was performed separately for each day based on the expected participant schedule and held fixed throughout that day. On the first day, the \AII{} malfunctioned: it repeated questions, and in one interview it produced long passages unrelated to the interview. We discarded all four \AII{}-led interviews from that day and revised the system before continuing (see System Architecture). The four human-led interviews from that day were randomized like those on other days and are retained, so 45 participants were randomized and 41 interviews are analyzed. Two later days had a single interview each. Appendix~\ref{app:robust} shows that no conclusion changes when conditions are compared only within days or when the first day is excluded. Participants were informed of their assignment upon arrival at the interview location.

The two conditions are balanced on every characteristic measured: gender, region of origin, degree level, field of study, prior ChatGPT use, and age (Appendix~\ref{app:regression}, Table~\ref{tab:balance}). With around twenty participants per condition, only a sizeable imbalance would be detectable.

\subsection*{Measures of Data Quality}

The central outcome of interest is data quality, assessed through three measures drawn from the qualitative methods literature \citep{SmallCalarco2022}.

\textit{Exposure.} \citet{SmallCalarco2022} identify exposure --- the time spent with participants --- as a necessary, though not sufficient, condition for high-quality qualitative evidence. Interviews were scheduled in slots of roughly half an hour, but the time actually spent varied considerably, so we measure exposure both as a volume and as a rate. The volume measures are (a)~\textit{mean response length}, the average number of characters per answer in an interview; (b)~\textit{total interview length}, the sum of all response characters; and (c)~the number of questions answered. Because these depend on how long an interview ran, we also report (d)~\textit{questions answered per minute} and (e)~\textit{characters produced per minute}, together with the duration itself, measured over the analyzed portion of each interview and so excluding time spent on the dropped third case. The rate measures ask how much material a given condition yields per unit of the researcher's time, which is the quantity at stake in the argument for scalability, and they do not require interviews to be of equal length.

\textit{Specificity.} The specificity of a response --- the degree to which it contains concrete, detailed accounts of particular experiences rather than abstract generalizations --- is widely regarded as a key indicator of qualitative data quality \citep{Gerson2020, Lareau2021, SmallCalarco2022, Weiss1995}. Specific responses allow researchers to grasp the particular situations and conditions in which respondents are embedded and to develop what \citeauthor{SmallCalarco2022} (\citeyear{SmallCalarco2022}:36) call cognitive empathy, which is ``the degree to which the researcher understands how those interviewed or observed view the world and themselves''. This type of understanding is a prerequisite for analytical judgment \citep{Lichterman2017}. Specificity was coded on a scale from 1 (\textit{very vague, no specific details}) to 5 (\textit{very specific, comprehensive and detailed information}), following the approach of \citet{Wuttke2024}; the full coding guidelines are provided in Appendix~\ref{app:coding}.

\textit{Relevance.} Responses must address the researcher's substantive concerns to generate useful data. Irrelevant responses --- arising from misunderstanding, distraction, or misinterpretation of the interview context --- are a recognized challenge to data quality \citep{Weiss1995}. Relevance was coded on a scale from 1 (\textit{completely off-topic}) to 5 (\textit{completely on-topic}).

Both specificity and relevance were coded by the first author and a student research assistant. Condition labels were removed before coding, although coders could in principle have inferred the condition from the style of the questions. Two classes of material are excluded from all measures. The first is the third substantive case described above, dropped because participants interviewed after the third day were never exposed to it. The second is a set of 22 messages in which the participant used the interface's skip button rather than answering; these are reported separately in the Results rather than treated as responses. Both exclusions are applied by locating the relevant scripted question in the transcript, so they do not depend on any reconstruction of which guide question a given answer belongs to.\footnote{Guide-section labels were generated automatically for the \AII{}-led interviews but reconstructed after the fact for the human-led ones, and we therefore avoid relying on them.} After these exclusions the interviews yield 909 question-answer pairs, 618 from \AII{}-led and 291 from human-led interviews. From the full set we drew a random sample of 400 for coding, balanced by condition; 395 of these fall outside the excluded third case, 200 from \AII{}-led interviews and 195 from human-led interviews.

Every coded question-answer pair was scored independently by both coders, and the score entering the analysis is the mean of the two. Inter-rater reliability is reported in Table~\ref{tab:irr}. For specificity, the two coders agreed exactly on 53 percent of pairs and to within one scale point on 94 percent of the pairs, giving an ordinal Krippendorff's $\alpha$ of 0.65 \citep{Krippendorff2012}. This falls just short of the conventional threshold of 0.667 for drawing tentative conclusions, and we treat the specificity results accordingly. Relevance shows the opposite pattern: exact agreement is 93 percent, but because almost every response was scored 5, there is little variance for a chance-corrected coefficient to work with and $\alpha$ falls to 0.37. We therefore read the relevance measure as establishing that responses were overwhelmingly on-topic in both conditions rather than as discriminating between them.

\begin{table}[H]
\centering
\caption{Inter-rater Reliability Between the Two Coders}
\label{tab:irr}
\begin{tabularx}{\textwidth}{Xrrrr}
\toprule
\textbf{Measure} & \textbf{Exact} & \textbf{Within 1} &
  \textbf{Krippendorff $\alpha$} & \textbf{Cohen $\kappa_w$} \\
\midrule
Specificity & 0.527 & 0.937 & 0.648 & 0.628 \\
Relevance   & 0.934 & 0.980 & 0.365 & 0.411 \\
\bottomrule
\end{tabularx}
\smallskip
\parbox{\textwidth}{\singlespacing\small \textit{Note:} Krippendorff's $\alpha$ uses the ordinal
difference function; Cohen's $\kappa_w$ is quadratically weighted.}
\end{table}

\subsection*{Statistical Analysis}

Because the outcome variables are not normally distributed, we compare the two groups with the Mann-Whitney U (MWU) test. All reported $p$-values are adjusted for the nine comparisons using Holm's step-down procedure \citep{Holm1979}.\footnote{Like the Bonferroni adjustment, Holm's procedure controls the family-wise error rate under arbitrary dependence among the tests, which matters here because several outcomes are algebraic functions of the same two quantities; unlike Bonferroni, it is uniformly more powerful. Table~\ref{tab:mwu} also reports Bonferroni-adjusted values, and no conclusion differs between the two.} We supplement these tests with OLS estimates of the condition difference, with and without fixed effects for the day of interview, since randomization was carried out within days (Appendix~\ref{app:regression}). Appendix~\ref{app:robust} repeats the comparisons within days and without the first day of data collection. We do not adjust for respondent characteristics; Appendix~\ref{app:regression} explains why and reports the sample composition in Table~\ref{tab:balance}.

Because two of our central claims are claims of similarity rather than of difference, we do not rest them on non-significance alone. A test that fails to reject at $N = 41$ may simply lack the power to detect a difference that matters. For each outcome we therefore also report the Hodges-Lehmann shift estimate --- the median of all between-group pairwise differences, the location parameter underlying the MWU test --- with a distribution-free confidence interval computed at the same family-wise level as the Bonferroni-corrected tests. The bound of that interval farthest from zero is the smallest difference our design is able to exclude: differences outside the interval are inconsistent with the data, while differences inside it cannot be ruled out.

% -------------------------------------------------------
\section*{Results}

\subsection*{Exposure}

\begin{figure}[H]
  \centering
  \includegraphics[width=0.88\linewidth]{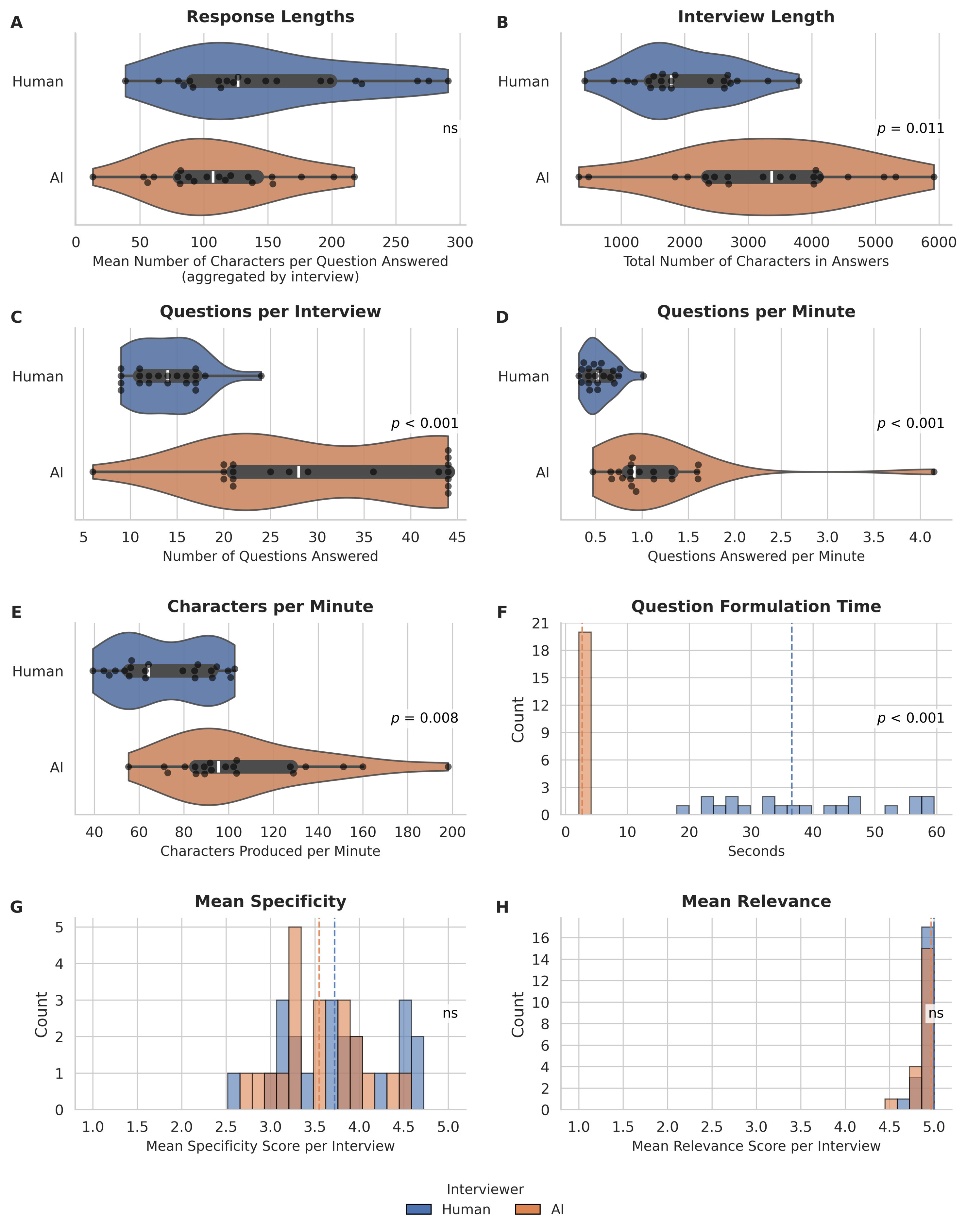}
  \caption{Human- and AI-led interviews compared. Exposure: (A) mean characters per answer, (B) total characters per interview, (C) questions answered, (D) questions answered per minute, (E) characters per minute, (F) mean question formulation time. Data quality: (G) mean specificity and (H) mean relevance per interview (scale 1--5). Violin plots (A--E) show the kernel density with the median (white dot), interquartile range (thick bar) and 1.5 IQR whiskers (thin line); histograms (F--H) mark each group's median with a dashed line. Labels give Holm-corrected Mann-Whitney U $p$-values; \textit{ns} = not significant.}
  \label{fig:main}
\end{figure}

Response lengths were relatively short overall, which is typical of written compared to verbal responses \citep{HohneGavras2022}. Human interviewers elicited longer individual responses --- averaging 150 characters ($SD = 72$) compared to 112 characters ($SD = 52$) for \AIIs{}, a 34 percent difference --- but this did not reach significance after correction for multiple testing (MWU $U = 147.0$, $p = 0.412$, Rank-Biserial Correlation $r = -0.300$; see Table~\ref{tab:mwu} and Figure~\ref{fig:main}). Human interviews also showed greater variability in response length, suggesting broader diversity in how much respondents elaborated per question.

When considering total interview length, the pattern reverses. AI-led interviews were significantly longer overall ($U = 327.5$, $p = 0.011$, $r = 0.560$, $\mathrm{CLES} = 0.780$), averaging 3,244 characters compared to 1,968 for human-led interviews --- a 65 percent difference. This is explained by the substantially greater number of questions asked: \AIIs{} averaged 31 questions compared to 14 for human interviewers ($U = 392.0$, $p < 0.001$, $r = 0.867$, $\mathrm{CLES} = 0.933$). The \AII{} figure is partly a design parameter: seven of the twenty \AII{}-led interviews reached the 44 questions the guide permits once every probe allowance is used, so the count reflects the probe budget set by the researcher as much as the behavior of the model. In written chat interviews, human interviewers required approximately 39 seconds to read a response and formulate a follow-up; the \AII{} averaged just 3 seconds ($U = 0.0$, $p < 0.001$, $r = -1.000$): the two distributions do not overlap at all.

The two conditions spent similar time with participants, though not identically so. The median interview ran 28.9 minutes under an \AII{} against 28.3 under a human interviewer ($U = 246.0$, $p = 0.769$). The means diverge further --- 30.3 against 26.7 minutes --- because five AI-led interviews ran past forty minutes where only one human-led interview did; the rank-based test is not driven by these few long interviews, which is why we report medians alongside it. We do not read this as establishing equal exposure time: the interval on the difference, at the same family-wise level as the other intervals, runs from $-4.6$ to $+12.8$ minutes, so an \AII{} advantage of up to nearly thirteen minutes is consistent with our data. The measures that follow are therefore reported as rates, which normalize by the duration of each interview and so hold whether or not that difference is real.

\AIIs{} obtained 1.18 answers per minute against 0.54 for human interviewers ($U = 387.0$, $p < 0.001$, $r = 0.843$) and 106 characters of respondent text per minute against 72 ($U = 333.0$, $p = 0.008$, $r = 0.586$). An \AII{} thus yields roughly 48 percent more material per minute of data collection, and more than twice as many answered questions per minute.

One feature of the \AII{} condition cuts the other way. The chat interface offered participants in both conditions a button to skip a question. It was used 22 times by five participants, all of them in \AII{}-led interviews; no participant in a human-led interview used it (5 of 20 against 0 of 21; Fisher's exact test, $p = 0.021$, uncorrected and outside the family of nine comparisons above). Skips are excluded from the answer counts and from mean response length, but not from duration, since they are time the interview spent without yielding an answer. Counting them instead as answers of zero length would lower mean response length in the \AII{} condition from 112 to 108 characters ($U = 141.0$, $p = 0.296$, Holm-adjusted), and no conclusion above changes. Counted as questions asked rather than answered --- since more questions asked means more ground covered --- they would raise the \AII{} rate only from 1.18 to 1.21 questions per minute.

\subsection*{Specificity and Relevance}

Neither specificity nor relevance differed significantly between conditions after correction for multiple testing (Table~\ref{tab:mwu}). Both conditions produced effectively identical mean relevance scores ($M = 4.9$ in both), with very little variation --- consistent with the deliberate nature of written responses, which tend to be more considered than verbal replies \citep{HohneGavras2022}. Specificity was marginally higher in human-led interviews ($M = 3.8$, $SD = 0.6$) than \AII{}-led interviews ($M = 3.6$, $SD = 0.5$), but this difference was not significant ($U = 166.0$, $p = 0.769$, $r = -0.210$).

Non-significance is not equivalence, so we ask what the design can actually exclude. For relevance the answer is decisive: the estimated shift is zero and the interval runs from $-0.11$ to $0.05$ scale points, so any difference in relevance larger than about a tenth of a point is inconsistent with our data. For specificity the answer is weaker. The estimated shift is $-0.16$ points in favor of human interviewers, with an interval from $-0.71$ to $0.36$. We can rule out a human advantage larger than roughly three quarters of a scale point (nearly a full point if the first day of data collection is excluded; Appendix~\ref{app:robust}), but a difference of half a point --- which on a five-point scale is not trivial --- remains possible with what we observe. The relevance result can therefore be read as equivalence on the measure as operationalized, with the caveat that, with scores clustered at the ceiling and modest inter-coder reliability, the measure registers only the coarse distinction between on- and off-topic answers; the specificity result is a failure to detect a difference.

Two qualifications apply. First, both conditions encountered difficulty eliciting developed, experientially grounded accounts, which may have limited the sensitivity of the specificity measure. Second, relevance was coded relative to the question asked and a brief project description; a stricter operationalization based on question-specific goals might produce more variation between conditions.

\begin{table}[H]
\centering
\caption{Mann-Whitney U Test Results ($N = 41$)}
\label{tab:mwu}
\begin{tabularx}{\textwidth}{Xrrrrr}
\toprule
\textbf{Outcome} & \textbf{$U$} & \textbf{$p$ (Holm)} &
  \textbf{$p$ (Bonf.)} & \textbf{RBC} & \textbf{CLES} \\
\midrule
Duration                  & 246.0 & 0.769    & 1.000    &    0.171 & 0.586 \\
Response length           & 147.0 & 0.412    & 0.928    & $-$0.300 & 0.350 \\
Interview length          & 327.5 & 0.011    & 0.020    &    0.560 & 0.780 \\
Questions per interview   & 392.0 & $<$0.001 & $<$0.001 &    0.867 & 0.933 \\
Questions per minute      & 387.0 & $<$0.001 & $<$0.001 &    0.843 & 0.921 \\
Characters per minute     & 333.0 & 0.008    & 0.013    &    0.586 & 0.793 \\
Question formulation time &   0.0 & $<$0.001 & $<$0.001 & $-$1.000 & 0.000 \\
Mean specificity          & 166.0 & 0.769    & 1.000    & $-$0.210 & 0.395 \\
Mean relevance            & 183.0 & 0.769    & 1.000    & $-$0.129 & 0.436 \\
\bottomrule
\end{tabularx}
\smallskip
\parbox{\textwidth}{\singlespacing\small \textit{Note:} Positive RBC values indicate \AII{} $>$ human-interviewer. RBC = Rank-Biserial Correlation; CLES = Common Language Effect Size.}
\end{table}

% -------------------------------------------------------
\section*{Discussion}

\subsection*{Scope and Interpretation}

Our results require care about what the comparison actually involves. We compare \AIIs{} to trained, non-expert human interviewers conducting short synchronous chat interviews --- not to the gold standard of qualitative interviewing. Expert qualitative interviewers, who have deeply internalized best practices and are well-versed in the relevant literature, can adjust interviews dynamically in response to emerging themes, deploy sophisticated listening and probing techniques, and build the trust and rapport central to eliciting candid, detailed accounts \citep{Weiss1995, SmallCalarco2022}. Our results should not be taken as evidence that \AIIs{} can replace expert human interviewers, who remain the appropriate choice wherever resources permit.

Our findings are instead relevant to the context of large-scale qualitative data collection, where non-expert research assistants --- typically with basic methodological training --- are rapidly prepared to administer a structured guide on a specific topic, as in projects such as the American Voices Project \citep{EdinFieldsEtAl_2024_Listening_to_Voices_of_America}. Within this context, our results suggest that \AIIs{} can elicit responses as on-topic as, and not detectably less specific than, those obtained by such interviewers, while offering substantially greater scalability.

At the same time, our results do not represent an upper bound on \AII{} performance. We used a relatively small model that was not fine-tuned for interview tasks and relied on general instruction prompts and simple classification filters. There is considerable room for improvement through task-specific fine-tuning, particularly as open-source datasets for training \AIIs{} become available.

The chat-based setting also constrains our comparison in ways that likely disadvantage human interviewers specifically. Chat interviews lack the social cues, vocal tone, and relational dynamics that facilitate trust and rapport in face-to-face settings \citep{dillman_survey_2005}. Our guide also omitted a rapport-building introduction --- a standard practice known to elicit longer and more detailed responses \citep{Gerson2020, Weiss1995} --- which likely had a greater negative effect on human than \AII{} interviewers. Future research should extend the comparison to face-to-face interviews to establish more ecologically valid benchmarks. We hypothesize that the human advantage in building rapport may be substantial in face-to-face contexts, where the constraints of the chat format are removed.

The concentration of skipped questions in the \AII{} condition is the clearest respect in which it did worse, and our design cannot say why. Three explanations are plausible. Declining a question may carry less social cost when the interviewer is a machine. Participants knew which condition they were in, and may have treated the \AII{} interview as closer to a survey. And the \AII{} often probed up to the limit the guide allowed, which may have made later probes feel repetitive. With five participants who skipped, we cannot distinguish between these, and we leave that question to future work.

\subsection*{Challenges in Measuring Qualitative Data Quality}

A second set of challenges concerns the quantification of qualitative standards. Unlike survey research, qualitative methodology lacks a consensus on standardized data quality measures. \citepos{SmallCalarco2022} recent attempt to articulate such criteria --- a valuable contribution --- has itself been contested \citep{Tavory2023}, underscoring the difficulty of the task. Experimental comparisons of qualitative interview methods therefore remain far less developed than in survey methodology.

Quantifying qualitative standards introduces difficulties familiar from the content analysis literature. First, \textit{semantic validity} \citep{Krippendorff2012}: do our operationalizations actually capture what we mean by data quality, or do they measure proxies that diverge from the underlying concept? Our measure of exposure uses character count as a proxy for depth, which assumes that longer responses yield richer material. This assumption is contestable: brief phrases can carry significant meaning within an interview context \citep{bamberg2008small, wooffitt2005conversation}, and length may reflect verbosity as much as depth. Second, \textit{unitization} \citep{Krippendorff2012}: we coded specificity and relevance using question-answer pairs as the unit of analysis. This ignores how meaning is shaped by conversational context --- a response cannot be fully interpreted in isolation from the exchange surrounding it. This fluidity in defining analytical units is at least partially incompatible with the demands of standardized quantitative comparison.

These challenges do not render the methodological agenda futile. Rather, they point to a productive frontier. The emergence of AI-interviewing creates both the need and the opportunity for more experimental approaches to qualitative methodology. The availability of large corpora of AI-generated interview transcripts, combined with the possibility of systematic manipulation of interview parameters, opens new directions for methodological research that were previously impractical.

\subsection*{Open-Weight Models and the Future of AI-Interviewing}

Our finding that a 7B-parameter open-weight model elicited responses as on-topic as, and not detectably less specific than, those obtained by non-expert human interviewers has practical implications for the methodological community. The GDPR constraints and infrastructure costs associated with large commercial APIs create significant barriers for many research groups, particularly those working with sensitive populations. Locally hosted open-weight models offer a viable alternative that keeps data under researcher control, supports reproducibility, reduces environmental impact, and removes dependence on commercial providers.

The capabilities of open-weight models are growing rapidly \citep{LuLiEtAl_2025_Small_Language_Models_Survey_Measurements_and_Insights}, and the performance gap with commercial systems is narrowing. The performance of LLMs on specific tasks depends substantially on the similarity of those tasks to their training data \citep{McCoyYaoEtAl_2024_Embers_of_autoregression_show_how_large_language_models_are_shaped_by_problem_they_are_trained_to_solve, McCoyYaoEtAl_2024_When_language_model_is_optimized_for_reasoning_does_it_still_show_embers_of_autoregression_analysis_of_OpenAI_o1}, suggesting that task-specific fine-tuning will yield especially large gains for interview applications. An open-source ecosystem for AI-interview development --- comprising shared datasets, standardized prompts, evaluation frameworks, and system architectures --- would accelerate this progress. We release our system prompts, interview guide, and analysis code at \url{https://github.com/ainterviewer/can-ai-interview} as a contribution toward this ecosystem.

While smaller open-weight models address many of the concerns associated with closed API systems, they are not fully transparent. A key limitation of models such as Mistral is the absence of documentation about their training data. This matters methodologically: the interactional, cultural, and knowledge competencies of an LLM derive largely from its training data, and undisclosed data introduce the risk of unexamined biases \citep{Navigli2023}. Just as human interviewers can systematically bias data through their characteristics and behaviors, so-called interviewer effects \citep{GrovesFultz1985, WestLi2019}, \AIIs{} may exhibit model-specific patterns that are difficult to detect without access to training data. The ideal \AII{} would be based on a fully open-source model with documented training data, architecture, and instruction sets. We experimented with OLMo 7B instruct, which satisfies these criteria, but were unable to achieve satisfactory interview performance on internal tests. We therefore used OpenHermes 2.5, which performed well enough and whose fine-tuning data are published, although the training data of the underlying Mistral 7B base model are not.

% -------------------------------------------------------
\section*{Conclusion}

This study provides the first direct empirical comparison of data quality in AI-led and human-led chat interviews. Within a non-expert, chat-based setting, \AIIs{} produce responses that are as consistently on-topic, and of specificity we cannot distinguish from that of trained human interviewers, while collecting substantially more material and more answered questions per minute of interviewing time. These findings support the viability of AI-interviewing as a method for large-$N$ qualitative data collection in sociological research.

Two broader lessons follow. First, the methodological infrastructure of AI-interviewing matters: locally hosted, open-weight language models address the data security, reproducibility, and ethical concerns associated with commercial API-based systems, and should be the default for social science applications wherever performance is adequate. Second, developing robust and widely shared measures of qualitative data quality --- a challenge that long predates AI-interviewing --- becomes increasingly urgent in AI-interviewing. AI-interview systems need to be evaluated on both ethical and epistemological grounds, and shared quality measures are essential if the method is to improve cumulatively rather than system by system. Meeting this challenge will require closer collaboration between qualitative methodologists and computational researchers.

% -------------------------------------------------------
\section*{Data Availability}

The materials underlying this study are publicly available, with the exceptions noted below, at the following GitHub repository: \url{https://github.com/ainterviewer/can-ai-interview}. The repository includes:
\begin{sloppypar}
\begin{itemize}
  \item \textbf{Aggregated data}: \texttt{data/agg.csv} (behavioral
    metrics) and \texttt{data/relevance\_and\_specificity\_scores.csv}
    (coded quality scores)
  \item \textbf{Coding data}: \texttt{data/coder\_agreement.csv}, holding
    both coders' independent scores for every coded pair, and
    \texttt{data/pair\_level\_scores.csv}, holding the coded scores
    together with the interview identifier for each coded answer.
    Neither file contains response text.
  \item \textbf{Interview instrument}: \texttt{data/interview\_guide.json},
    containing the full semi-structured interview guide including
    framing, introductory and closing texts, all main questions,
    follow-up probes, and conditional skip logic
  \item \textbf{AI system prompts}: Jinja prompt templates for the
    probing, reformulation and classification agents in the
    \texttt{prompts/} directory
  \item \textbf{Design data}: \texttt{data/interview\_day.csv}, mapping each
    interview to the day of data collection as an index carrying no
    calendar date
  \item \textbf{Analysis code}: \texttt{analysis.ipynb} with all
    statistical analyses, along with \texttt{requirements.txt}
  \item \textbf{Derivation and checking code}: \texttt{scripts/build\_data.py},
    which documents how every data file above was produced from the confidential message-level
    transcripts and the two coders' score sheets, applying the
    exclusions described in the Measures section;
    \texttt{scripts/stats\_tables.py}, which reproduces the comparisons in
    Table~\ref{tab:mwu}, the equivalence bounds and the robustness
    checks in Table~\ref{tab:robust}; and
    \texttt{scripts/verify\_paper.py}, which recomputes every quantity
    reported in this article and checks it against the manuscript source
\end{itemize}
\end{sloppypar}
Raw interview transcripts are not shared in order to protect participant confidentiality. Interview responses were collected under a confidentiality agreement that precludes sharing verbatim text. Respondent-level demographic characteristics are withheld for the same reason: in a sample of this size the combination of age, gender, region of origin, study programme and degree level is close to uniquely identifying, and the repository publishes an interview identifier on which it could be joined. These characteristics are reported in aggregate in Table~\ref{tab:balance}.

% -------------------------------------------------------
\section*{Ethics Declaration}

Ethical approval was granted by the institutional review board of the corresponding author's university. All research was conducted in accordance with relevant institutional guidelines and regulations. Informed consent was obtained from all participants prior to participation. Participants were informed of the purpose of the study, the voluntary nature of their participation, and that all responses would be treated confidentially with personally identifiable information removed from publications.

% -------------------------------------------------------
\bibliographystyle{asa}
\bibliography{references}

% -------------------------------------------------------
\newpage
\appendix

\section{Coding Guidelines}
\label{app:coding}

A random sample of 400 question-answer pairs was coded for specificity and relevance by the first author and a student research assistant; 395 of these (200 from \AII{}-led and 195 from human-led interviews) fall outside the excluded third case and enter the analysis. Every sampled pair was scored independently by both coders and the two scores averaged; the resulting pair-level scores were then averaged within each interview to yield per-interview mean scores.

\subsection*{Specificity}

Evaluate the level of specificity and detail in the response. Specific responses provide sufficient detail to allow not only understanding but also the development of in-depth analytical insights.

\begin{itemize}
  \item[1:] Very vague, with no specific details.
  \item[2:] Mostly vague, with few specific details.
  \item[3:] Somewhat specific, with some detailed information.
  \item[4:] Mostly specific, with substantial detailed information.
  \item[5:] Very specific, with comprehensive and detailed information.
\end{itemize}

\subsection*{Relevance}

Evaluate how well the response aligns with the topic or question asked. Relevant responses indicate respondent engagement with the substance of the question.

\begin{itemize}
  \item[1:] Response is completely off-topic.
  \item[2:] Response is mostly off-topic.
  \item[3:] Response is somewhat relevant but includes off-topic
    information.
  \item[4:] Response is mostly relevant to the topic.
  \item[5:] Response is completely relevant to the topic.
\end{itemize}

% -------------------------------------------------------
\section{Sample Composition and Regression Results}
\label{app:regression}

Table~\ref{tab:balance} reports the composition of the two conditions, and motivates the specifications that follow: because the groups are balanced on every characteristic measured, adjusting the estimates for those characteristics buys little, while at $N = 41$ it costs a great deal of precision.

\begin{table}[H]
\centering
\caption{Sample Composition by Condition}
\label{tab:balance}
\begin{tabularx}{\textwidth}{Xrrr}
\toprule
 & \textbf{AI} & \textbf{Human} & \textbf{$p$} \\
\midrule
\textit{Gender} & & & 1.00 \\
\quad Female & 10 & 12 & \\
\quad Male & 9 & 9 & \\
\addlinespace
\textit{Region of origin} & & & 0.51 \\
\quad Nordic & 11 & 16 & \\
\quad Other EU & 2 & 1 & \\
\quad Non-EU & 6 & 4 & \\
\addlinespace
\textit{Degree level} & & & 0.52 \\
\quad Bachelor & 10 & 14 & \\
\quad Master & 9 & 7 & \\
\addlinespace
\textit{Field of study} & & & 0.24 \\
\quad Data and computational & 8 & 11 & \\
\quad Economics & 6 & 2 & \\
\quad Other social science & 5 & 8 & \\
\addlinespace
\textit{ChatGPT use} & & & 0.76 \\
\quad Several times a day & 5 & 6 & \\
\quad About once a day & 4 & 4 & \\
\quad A few times a week & 7 & 5 & \\
\quad Once a week or less & 3 & 6 & \\
\addlinespace
\textit{Age}, mean (SD) & 24.8 (3.0) & 25.0 (5.3) & 0.36 \\
\midrule
\textit{N} & 19 & 21 & \\
\bottomrule
\end{tabularx}
\smallskip
\parbox{\textwidth}{\singlespacing\small \textit{Note:} One AI-led interview is omitted because the
participant did not complete the post-interview survey, so $N = 40$ here
against $N = 41$ elsewhere. $p$-values are from Fisher's exact test for binary
characteristics, a Monte Carlo permutation $\chi^{2}$ test for the remainder,
and a Mann-Whitney U test for age. Study programmes are grouped into fields;
ungrouped programme counts are not reported because several contain a single
participant.}
\end{table}

Table~\ref{tab:reg} reports OLS estimates of the condition difference for each interview-level outcome. Model~1 includes the condition indicator alone; Model~2 adds fixed effects for the day of interview. Because randomization was carried out separately within each day and the share of participants assigned to each condition varied between days, Model~2 is the estimate the design supports, and Model~1 is reported for comparison. Days on which only one condition is represented, including the first day of data collection, contribute nothing to the Model~2 coefficient, which is therefore identified from the 35 interviews on the remaining seven days. The \AII{} coefficient is the estimated difference relative to the human-interviewer condition.

The estimates corroborate the non-parametric findings. \AIIs{} produce longer total interviews and ask substantially more questions; respondents give shorter individual answers to \AIIs{}, though that coefficient reaches conventional significance only under day fixed effects ($-40.25$, $p = 0.037$) and not without them ($-38.14$, $p = 0.061$). These $p$-values are not adjusted for multiple testing, and the Holm-adjusted rank tests find no significant difference in response length under any of the specifications in Table~\ref{tab:robust}. Neither specificity nor relevance differs significantly in either model. The specificity coefficient is larger under day fixed effects ($-0.290$, $p = 0.102$) than without them ($-0.200$, $p = 0.257$), but remains short of conventional significance.

We do not adjust for respondent characteristics. Under randomization such adjustment is not required for identification, and with $N = 41$ a specification carrying dummies for age, gender, region, programme, degree level and prior ChatGPT use would retain roughly a dozen residual degrees of freedom --- too few to support the coefficients it produces. Sample composition is reported in Table~\ref{tab:balance}. Heteroskedasticity-consistent (HC3) standard errors are reported for Model~1, where the two conditions differ markedly in variance; they are not defined for Model~2, because two of the ten days contain a single interview whose day fixed effect fits it exactly.

\begin{table}[H]
\centering
\caption{OLS Regression: \AII{} Condition (Human = Reference), $N = 41$}
\label{tab:reg}
\begin{tabularx}{\textwidth}{Xrrrrr}
\toprule
 & \textbf{Interview} & \textbf{Response} & \textbf{Question} &
   \textbf{Specificity} & \textbf{Relevance} \\
 & \textbf{length} & \textbf{length} & \textbf{count} & & \\
\midrule
\multicolumn{6}{l}{\textit{Model 1: condition only}} \\
\quad \AII{} &
  $1275.77^{**}$ & $-38.14$ & $17.04^{***}$ & $-0.200$ & $-0.031$ \\
 & $(374.64)$ & $(19.75)$ & $(2.78)$ & $(0.174)$ & $(0.032)$ \\
\quad HC3 s.e.
 & $[389.29]$ & $[20.08]$ & $[2.91]$ & $[0.177]$ & $[0.033]$ \\
\quad $R^{2}$ & 0.229 & 0.087 & 0.490 & 0.033 & 0.022 \\
\quad Adj. $R^{2}$ & 0.209 & 0.064 & 0.477 & 0.008 & $-0.003$ \\
\addlinespace
\multicolumn{6}{l}{\textit{Model 2: condition and day fixed effects}} \\
\quad \AII{} &
  $1052.01^{**}$ & $-40.25^{*}$ & $15.34^{***}$ & $-0.290$ & $-0.042$ \\
 & $(315.15)$ & $(18.39)$ & $(2.98)$ & $(0.172)$ & $(0.040)$ \\
\quad $R^{2}$ & 0.674 & 0.527 & 0.650 & 0.435 & 0.091 \\
\quad Adj. $R^{2}$ & 0.566 & 0.370 & 0.534 & 0.246 & $-0.212$ \\
\bottomrule
\end{tabularx}
\smallskip
\parbox{\textwidth}{\singlespacing\small \textit{Note:} Classical standard errors in parentheses,
HC3 standard errors in brackets. Model~1 has 39 residual degrees of
freedom, Model~2 has 30. $^{*}p < 0.05$, $^{**}p < 0.01$,
$^{***}p < 0.001$.}
\end{table}

% -------------------------------------------------------
\section{Robustness to the Day Structure}
\label{app:robust}

Randomization was carried out within days, but not every day contributes both conditions: on the first day of data collection only the human-led interviews were retained, and two later days had a single interview each. The share of participants assigned to the \AII{} condition therefore varied between days, and the pooled comparisons in Table~\ref{tab:mwu} could in principle be confounded by day. The first day matters most. Its four human-led interviews produced shorter answers (98 against 162 characters on average) and lower mean specificity (3.21 against 3.88) than the 17 human-led interviews on later days, so including them lowers the human-led average on exactly the measures where the conditions are compared for similarity.

Table~\ref{tab:robust} repeats the nine comparisons in two ways. The first compares the conditions within days only, using the van Elteren test, the stratified counterpart of the Mann-Whitney U test \citep{vanElteren1960}; days on which only one condition was run do not contribute. The second excludes the first day and repeats the Mann-Whitney tests on the remaining 37 interviews. As in the main text, all $p$-values are Holm-adjusted across the nine comparisons.

\begin{table}[H]
\centering
\caption{Holm-adjusted $p$-values Under Alternative Treatments of the Day Structure}
\label{tab:robust}
\begin{tabularx}{\textwidth}{Xrrr}
\toprule
\textbf{Outcome} & \textbf{Full sample} & \textbf{Within days} &
  \textbf{Without day 1} \\
\midrule
Duration                  & 0.769    & 1.000    & 0.794    \\
Response length           & 0.412    & 0.249    & 0.159    \\
Interview length          & 0.011    & 0.016    & 0.028    \\
Questions per interview   & $<$0.001 & 0.001    & $<$0.001 \\
Questions per minute      & $<$0.001 & $<$0.001 & $<$0.001 \\
Characters per minute     & 0.008    & 0.016    & 0.026    \\
Question formulation time & $<$0.001 & $<$0.001 & $<$0.001 \\
Mean specificity          & 0.769    & 0.530    & 0.246    \\
Mean relevance            & 0.769    & 1.000    & 0.794    \\
\bottomrule
\end{tabularx}
\smallskip
\parbox{\textwidth}{\singlespacing\small \textit{Note:} \textit{Full sample} repeats the Holm-adjusted column of Table~\ref{tab:mwu}. \textit{Within days}: van Elteren test stratified by day of data collection ($N = 41$, of which the 35 interviews on days with both conditions contribute). \textit{Without day 1}: Mann-Whitney U test with the first day excluded ($N = 37$).}
\end{table}

No conclusion changes. The exposure differences remain significant under both alternatives, and duration, response length, specificity and relevance remain non-significant. The equivalence bounds for specificity do become less favorable to the \AII{}. Without the first day, the Hodges-Lehmann interval for the difference in specificity runs from $-0.94$ to $0.21$, against $-0.71$ to $0.36$ in the full sample, so a human advantage of almost a full scale point can no longer be excluded. The interval for relevance, from $-0.11$ to $0.03$, is essentially unchanged. The regression estimates point the same way: Model~2 in Table~\ref{tab:reg}, which draws only on days on which both conditions were run, gives a specificity coefficient of $-0.290$, against $-0.200$ without day fixed effects.

% -------------------------------------------------------
\section{Interview Guide}
\label{app:guide}

The full interview guide in machine-readable format (\texttt{interview\_guide.json}) is available in the data repository. The guide is reproduced here in abbreviated form to document the study instrument.

\subsection*{Framing (for model use only)}

This interview investigates university students' academic help-seeking behavior through two cases: (1) a situation in which the student needed and received help, and (2) a situation in which the student needed help but did not ask for it. The aim is to investigate social processes and mechanisms surrounding help-seeking --- why students need help, who they seek it from, the social context, and students' feelings, attitudes, and behaviors.

\subsection*{Introduction (shown to participants)}

Thank you very much for taking part in our research. In this interview, we will ask you open-ended questions to investigate university students' academic help-seeking behavior. There are no right or wrong answers. We are very interested in your experience and perspective. The only constraint is that you write your answers in one go, rather than in multiple messages. Your responses will remain confidential and all personally identifiable information will be removed from publications.

\subsection*{Section 1: Receiving Help}

\begin{enumerate}
  \item Can you think of a situation where you needed academic help and
    got help with your studies? (Yes/No)
  \item What did you specifically need help with?
  \item How did you get help? (\textit{Probes, max 5}: Did you
    personally ask? From whom? How would you describe your relation to
    them? How did you decide who could help best and how to approach
    them?)
  \item Who else did you consider asking? (\textit{Probe, max 2}: What
    factors influenced your decision?)
  \item How did you feel after getting help? (\textit{Probe, max 2}:
    What are your expectations of the helper?)
  \item In general, what are the reasons you seek academic help?
    (\textit{Max 1 probe})
  \item Who do you ask to resolve academic challenges? How would you
    describe your relation to them? (\textit{Max 1 probe})
\end{enumerate}

\subsection*{Section 2: Not Asking for Help}

\begin{enumerate}
  \item Can you think of a situation where you needed help but did not
    ask for it? (Yes/No)
  \item What did you specifically need help with?
  \item Although you did not ask, did you consider asking anyone?
    (\textit{Probes, max 3}: Relation to them? Considerations around
    asking?)
  \item Why did you not ask for help? (\textit{Probes, max 4}: When do
    you feel most/least comfortable asking for help?)
  \item How did you resolve the issue? (\textit{Probes, max 5}: Did
    you end up getting help? Who could help best and how would you
    approach them?)
  \item How did you feel about this situation? (\textit{Probes, max 4}:
    Under what conditions would you seek help? What changes at the
    university would better meet your needs?)
  \item In general, in which cases or from whom do you refrain from
    asking for academic help? Why? (\textit{Max 1 probe})
\end{enumerate}

\subsection*{Closing}

Before we conclude, is there anything you would like to add or any perspectives we may have missed?

\end{document}